\documentclass[journal,twoside,web]{ieeecolor}
\usepackage{tmi}
\usepackage{cite}
\usepackage{amsmath,amssymb,amsfonts}
\usepackage{algorithmic}
\usepackage{graphicx}
\usepackage{stfloats}
\usepackage{textcomp}
\usepackage{arydshln}
\usepackage{hyperref}
\hypersetup{
    colorlinks = true,
    linkcolor = blue,
    anchorcolor = blue,
    citecolor = blue,
    filecolor = blue,
    urlcolor = blue
}

\begin{document}
\title{Synthetic PET via Domain Translation of 3D MRI}
\author{Abhejit~Rajagopal, Yutaka~Natsuaki, Kristen~Wangerin, Mahdjoub~Hamdi, Hongyu~An, John~J.~Sunderland, Richard Laforest, Paul~E.~Kinahan, Peder~E.Z.~Larson, Thomas~A.~Hope
\vspace{-12mm}
\thanks{This work involved human subjects or animals in its research. Approval of all ethical and experimental procedures and protocols was granted by the UCSF IRB under RB 17-21852.
Author affiliations: A.~Rajagopal, P.E.Z.~Larson, and T.A.~Hope are with the Department of Radiology and Biomedical Imaging, University of California, San Francisco. Y.~Natsuaki is with the Department of Radiation Oncology, University of New Mexico, Albuquerque.
K.~Wangerin is with G.E.~Healthcare, Waukesha, Wisconsin.
M.~Hamdi, R.~Laforest, and H.~An are with the Department of Radiology, Washington University, St. Louis. J.J~Sunderland is with the Department of Radiology, University of Iowa. P.E.~Kinahan is with the Department of Radiology, University of Washington, Seattle. Correspondence email: abhejit.rajagopal@ucsf.edu.}}
\markboth{Submitted to IEEE Transactions on Radiation \& Plasma Medical Sciences, 2021}
{Rajagopal \MakeLowercase{\textit{et al.}}: Synthetic PET via Domain Translation of 3D MRI}
\maketitle
\begin{abstract}
Historically, patient datasets have been used to develop and validate various reconstruction algorithms for PET/MRI and PET/CT. To enable such algorithm development, without the need for acquiring hundreds of patient exams, in this paper we demonstrate a deep learning technique to generate synthetic but realistic whole-body PET sinograms from abundantly-available whole-body MRI. Specifically, we use a dataset of 56 $^{18}$F-FDG-PET/MRI exams to train a 3D residual UNet to predict physiologic PET uptake from whole-body T1-weighted MRI. In training we implemented a balanced loss function to generate realistic uptake across a large dynamic range and computed losses along tomographic lines of response to mimic the PET acquisition. The predicted PET images are forward projected to produce synthetic PET time-of-flight (ToF) sinograms that can be used with vendor-provided PET reconstruction algorithms, including using CT-based attenuation correction (CTAC) and MR-based attenuation correction (MRAC). The resulting synthetic data recapitulates physiologic $^{18}$F-FDG uptake, e.g. high uptake localized to the brain and bladder, as well as uptake in liver, kidneys, heart and muscle. To simulate abnormalities with high uptake, we also insert synthetic lesions. We demonstrate that this synthetic PET data can be used interchangeably with real PET data for the PET quantification task of comparing CT and MR-based attenuation correction methods, achieving $\leq 7.6\%$ error in mean-SUV compared to using real data. These results together show that the proposed synthetic PET data pipeline can be reasonably used for development, evaluation, and validation of PET/MRI reconstruction methods.
\end{abstract}
\begin{IEEEkeywords}
PET/MRI qualification, digital phantoms, full-convolutional neural networks, SUV quantification
\end{IEEEkeywords}
\IEEEpeerreviewmaketitle

\newcommand{\tildee}{{\raise.17ex\hbox{$\scriptstyle\mathtt{\sim}$}}}

\section{Introduction}
\IEEEPARstart{T}{here} is currently an unrealized potential for PET/MRI systems in synergistic and quantitative reconstructions that account for and leverage simultaneous data acquisition of PET, which provides functional tissue information, and MRI, which provides excellent anatomic information, to correct for artifacts, motion, and improve localization~\cite{hope2019summary}. 
An example of one of the challenges is quantitative PET reconstruction, which requires accurate attenuation maps that are not directly measured by MRI.
This affects quantification of PET from reconstructed imagery, since the photon attenuation map is embedded in the forward system model.
{
As a result, 
}
the development of novel attenuation correction methods and other advanced PET/MRI reconstructions requires real or realistic data, which can be difficult and/or expensive to obtain.

{
With increased PET/MRI
}
adoption, it is necessary to establish standards for the quality of reconstruction, which can vary based on subtleties of PET data collection including scanner geometry and detector non-idealities, but also the choice of reconstruction algorithm, attenuation correction method, and patient anatomy (e.g.~scattering, hyper-attenuation).
{
Simulating the whole range of patient variability in terms of anatomy and patient-specific radiotracer uptake is infeasible,
e.g. using purely digital phantoms and Monte Carlo
simulation~\cite{abadi2020virtual,paredes2021simpet},
necessitating the acquisition of real patient 
}
PET data. For PET/CT systems, qualification methods are established by the American College of Radiology (ACR) using qualitative evaluation of whole-body clinical scans and quantitative evaluation using a ACR PET Phantom, a cylinder based on the Jaszczak Deluxe Flangeless ECT phantom with the spheres removed, a PET faceplate composed of several fillable cylinders, and acrylic rods of various diameters~\cite{macfarlane2006acr}. PET reconstruction performance can also be
{
measured
}
using the NEMA phantom, which is composed of multiple fillable spheres and cylindrical inserts that aim to mimic attenuation and activity found in different parts of the human body~\cite{ziegler2015nema}.

Unfortunately, phantoms used for PET/CT are insufficient for evaluating PET/MRI reconstruction performance because they cannot evaluate modern MR-based attenuation correction (MRAC) methods that rely on detecting typical human anatomy from MRI data.
These methods include vision-based atlas techniques~\cite{wollenweber2013evaluation,wollenweber2013comparison,torrado2016fast}, joint-reconstruction of attenuation and activity~\cite{defrise2013simultaneous,rezaei2018joint}, and direct prediction of pseudo-CT via deep learning-based domain translation~\cite{leynes2018zero,gong2018attenuation,leynes2020bayesian,chen2021deep}. 
A physical PET/MRI phantom to evaluate reconstruction performance would require an anthropomorphic distribution of materials with properties that match both 511 keV photon attenuation as well as MRI properties of proton density, T1, and T2, which is extremely challenging especially for bone due to its high attenuation but rapid T2 decay rate~\cite{chandramohan2020bone}.

Consequently, the standard approach to evaluating PET/MRI performance is to utilize human subject datasets acquired on PET, CT, and MR.~\cite{leynes2018zero,catana2021path}.
This allows for a \textit{relative} performance measure, by comparing the standardized uptake value (SUV) of MRAC-based PET reconstructions relative to PET reconstructions utilizing CT-based attenuation correction (CTAC)~\cite{catana2021path}. However, for sites to conduct such evaluations, numerous PET/CT/MR scans are required to characterize scanner and algorithm performance at operating points exhibiting natural imperfections that impact the physics of PET collection, such as those arising from detector characteristics, scattering, or unexpected attenuation~\cite{laforest2020harmonization}. This patient-specific data is expensive to collect, hindering new PET/MRI algorithm development that normally requires re-collecting PET data.

\begin{figure*}[!bt]
    \centering
    \includegraphics[width=\linewidth]{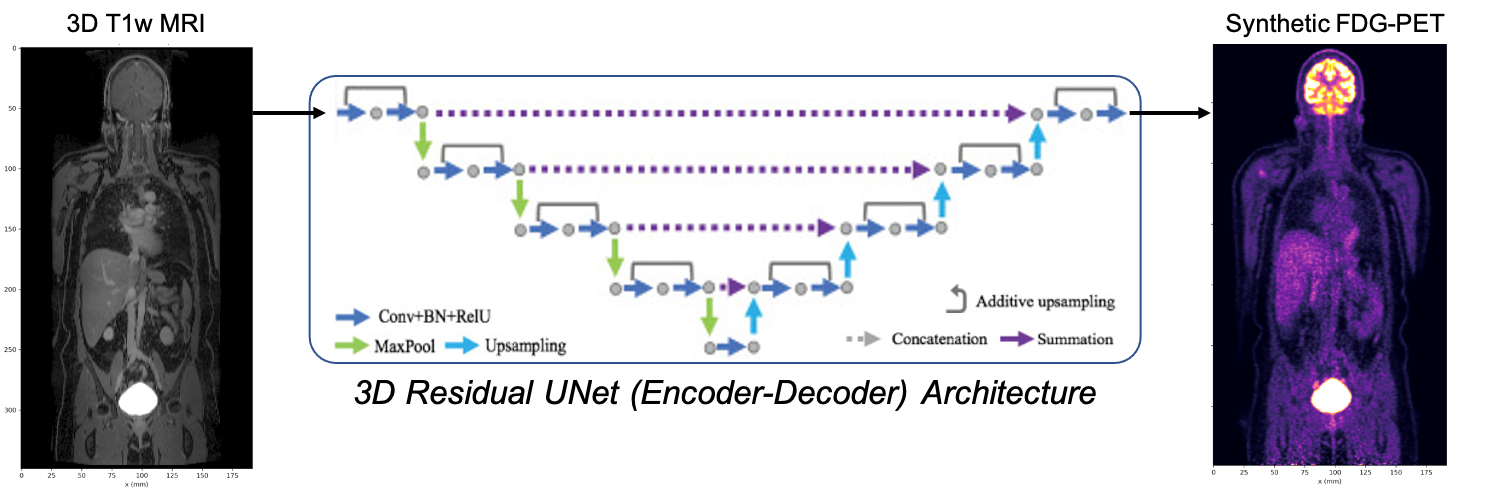}
    \vspace{-6mm}
    \caption{3D residual UNet architecture for generating synthetic PET from MRI, requiring only paired (registered) PET/MRI data without annotation.}
    \label{fig:architecture}
\end{figure*}

In this paper we present a method for generating \textit{synthetic} PET data using routinely-collected and abundantly available MRI that naturally captures important scanner and detector imperfections, adapts to varied tracer distributions and anatomy, and allows for insertion of synthetic lesions.
We believe this will allow for creation of large and diverse synthetic data for development, evaluation, and validation of PET reconstruction algorithms.
Our approach leverages recent work in deep learning-based domain translation using fully-convolutional networks (FCNs) and in Section~\ref{sec:domaintranslation} we describe how to train a 3D residual UNet to predict SUV-normalized synthetic PET imagery from whole-body postconstrast T1-weighted MRI (Fig.~\ref{fig:architecture}). 
This requires only paired input and output image examples, and---crucially---no additional annotation or scanner geometry details.
For this problem, we assume a supervised setting, where the absolute and relative error between the measured (reconstructed) and FCN-generated volumes provide a quantitative measure of performance, albeit at different scales that must be balanced.
{
Note that an approach based purely on generative adversarial networks (GANs) is not desirable here, since we require the synthetic PET volumes to correspond anatomically to the MRI volumes to support PET/MRI reconstruction research. To this end, in
}
Section~\ref{sec:quantification}, we show that the predicted synthetic PET (sPET) imagery can be forward projected to generate synthetic PET time-of-flight (TOF) sinogram data that can be used interchangeably with real PET sinogram data in vendor-provided reconstruction algorithms. We further evaluate this capability for qualification research by performing a classical PET-SUV quantification experiment, comparing reconstructions with CT- and 2-point MR-Dixon-based AC maps, using both synthetic FDG-PET and measured FDG-PET sinograms. Our results show that evaluation using synthetic PET can achieve $< 8\%$ quantification error in mean-SUV in synthetically-inserted lesions compared real PET data (averaged over a several synthetic lesions in a cohort of patients), suggesting the wide applicability of domain-translated synthetic PET for PET reconstruction algorithm development and qualification research. 
The role of synthetic lesions as proposed and demonstrated in this study is to provide methods for evaluation and optimization of image reconstruction algorithms. These algorithms continue to change and, with the introduction of deep learning/ AI methods for image reconstruction and denoising algorithms, many new parameters are being introduced and more robust methods for evaluation and optimization are needed to demonstrate the clinical impact of the image processing algorithms

\subsection{Prior Work}
Prior work in deep learning-based domain translation has demonstrated that FCNs based on UNet-like encode-decoder architectures are widely applicable to a range of
{
2D and 3D cross-modality medical image translation tasks,
including MRI-to-CT~\cite{leynes2018zero}, PET-to-CT~\cite{armanious2019unsupervised}, and MRI-to-MRI\cite{yang2020mri}. These architectures are typically trained independently for each anatomical region (e.g.~head, chest, pelvis) of interest. For PET/MRI specifically, a major focus has been in MRI-to-CT domain translation for enhanced attenuation correction maps, which are ultimately combined with measured PET sinogram data for enhanced image reconstruction~\cite{leynes2018zero,armanious2019unsupervised,leynes2020bayesian}.
}

{
Recently, such architectures have been applied to the reconstruction and denoising of low-dose PET imagery, including using supervised~\cite{ouyang2019ultra,xu2020ultra} and unsupervised~\cite{gong2018pet} methods, and extensions to dynamic PET reconstruction~\cite{yokota2019dynamic}.
In some cases, these image-enhancement techniques have been shown to successfully improve diagnostic interpretability~\cite{zaharchuk2020ai}. 
}

{
In contrast to these works, the focus of this article is domain translation of whole-body MRI-to-PET \textit{without} any initial PET data, i.e.~to produce a novel image series we refer to as synthetic~PET (sPET). While previously in-silico PET image generation has been explored using physics-based simulation tools such as GATE~\cite{jan2004gate} with Monte Carlo techniques like PENELOPE~\cite{espana2009penelopet} and SimPET~\cite{paredes2021simpet} to reproduce realistic image quality, a predominant issue here is knowing realistic spatial distribution of physiologic PET uptake to seed the simulation. Our work addresses this issue by using a neural network to learn from real PET scans, such that realistic physiologic uptake can be inferred from abundantly available MRI.
This is an important point since we do not believe sPET can accurately predict patient-specific functional information.
}

\subsection{Contributions}
Thus, our contributions are as follows:
\begin{itemize}
    \item We introduce a deep learning method for generating whole-body 3D synthetic PET (sPET) volumes from one or more routinely collected MRI series, including a balanced loss function that improves reconstruction of both low- and high-SUV regions.
    \item We evaluate the utility of sPET in a downstream development task involving the quantification of PET SUV in images reconstructed using MR- and CT-based attenuation correction, demonstrating that sPET sinograms can be used seamlessly in place of real PET data for PET/MRI qualification
    with minimal impact to the observed quantification error in synthetically-inserted lesions.
\end{itemize}

\section{Synthetic PET via Domain Translation} \label{sec:domaintranslation}
Although the physics and acquisition are fundamentally different, MRI and PET imagery share a great deal of structural similarity due to contouring of patient anatomy by physiologic uptake. This similarity can be exploited by FCNs to efficiently and \textit{implicitly} implement the codebook $\mathcal{C} : \mathbb{R}^n_\text{MR} \rightarrow \mathbb{R}^n_\text{PET}$ mapping MRI to PET-SUV imagery using a cascade of nonlinear filters, avoiding explicit storage of input-output pairs ($x$,$y$) in a database. 
{
Note that this map $\mathcal{C}$ describes a \textit{statistical} relationship between MRI and PET, and not a causual or functional one.
}
Besides being differentiable and amenable to backpropagation-based optimization using historical PET/MRI datasets, FCNs have strong spatial regularization properties that reduce degeneracy across image patches to create seamless and realistic anatomy-conforming 3D PET imagery from MRI.

Here, degeneracy refers to the typical inconsistencies in the codebook arising from the fact that PET and MRI contain different (orthogonal) information about a patient. The inverse image $\mathcal{C}^{-1}(y)$ of a 3D PET patch $y \in \mathbb{Z}^n_\text{PET}$ may not be unique, since different anatomical regions can experience the same uptake. Conversely, a given 3D MR patch $x \in \mathbb{R}^n_\text{MR}$ may have multiple images $\mathcal{C}(x) \in \mathbb{Z}^n_\text{PET}$, corresponding to various patterns of PET uptake across individuals. Thus the map $\mathcal{C}$ is general, which frustrates conventional atlas and dictionary-based implementations that must keep track of this in $\mathbb{R}^n_\text{MR}$~\cite{zhang2015beyond}.
In comparison, due to the supervised training process, FCNs naturally choose $y \leftarrow \mathop{{}\mathbb{E}}[\mathcal{C}(x)]$ for sPET given input MRI $x$. In this respect, the task of predicting PET from MRI is distinct from approaches predicting full-dose PET imagery from low-dose PET imagery, since those models are responsible for enhancing the SNR of existing activity images~\cite{ouyang2019ultra,xu2020ultra},
{
rather than directly learning anatomy-conforming physiologic biodistributions of PET uptake.
}

\begin{figure*}[tp]
    \centering
    \includegraphics[width=0.98\linewidth]{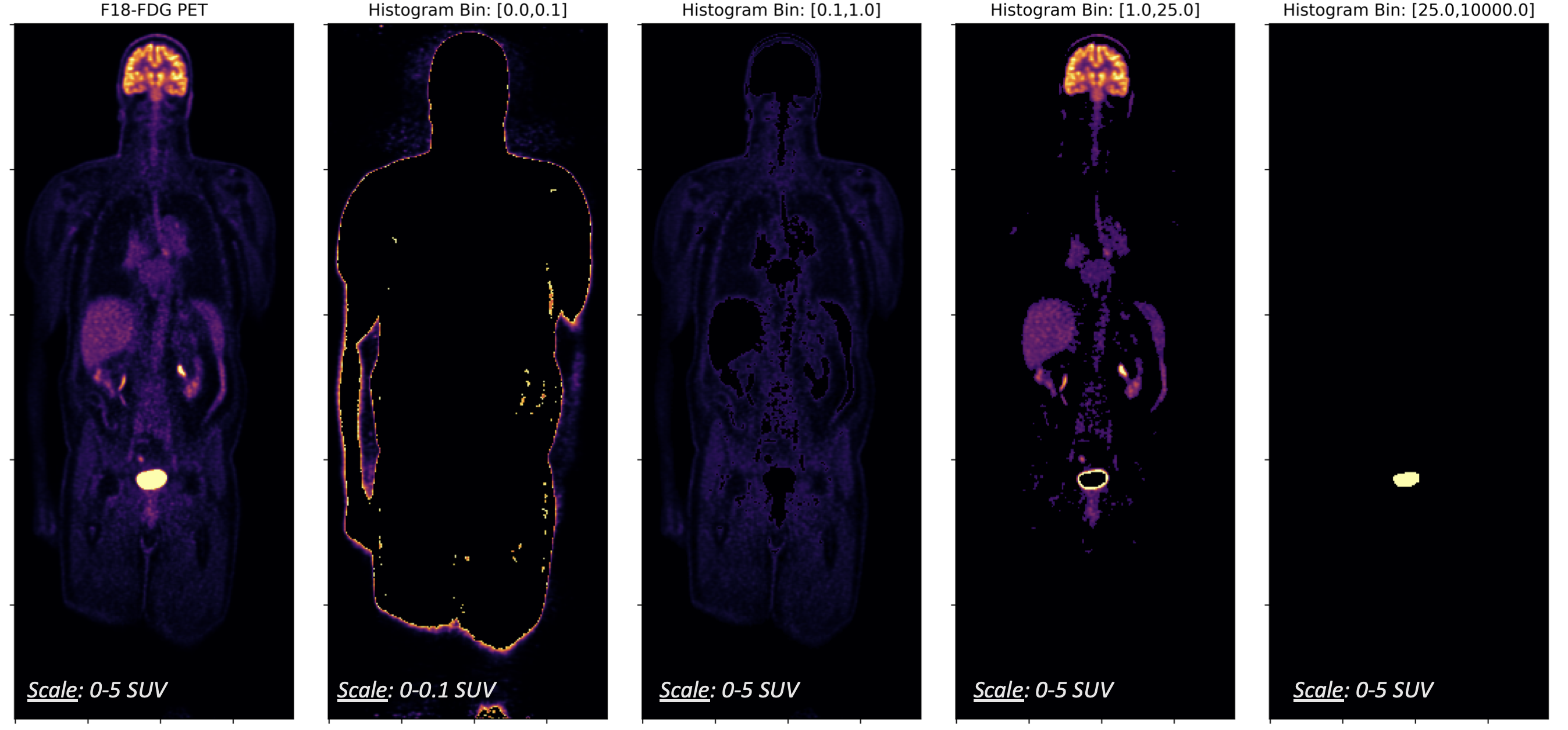}
    \caption{The histogram distribution of a whole-body PET exam reveals disparate levels of physiologic activity across different anatomy.}
    \label{fig:PEThisto}
\end{figure*}

\subsection{Assumptions} \label{subsec:assumptions}
In this paper, we assume the availability of historical PET/MRI datasets of patients receiving a calibrated (full) dose of the same PET radiotracer. Although the proposed method is applicable to varying dose levels, low-dose PET imagery exhibits lower signal-to-noise ratio (SNR), and is therefore not ideal for training. In this work, we register scanner-reconstructed whole-body $^{18}$F-FDG-PET and post-contrast T1-weighted MRI volumes, collected on a 3.0~T time-of-flight (ToF) PET/MRI scanner (Signa, GE Healthcare, Waukesha WI), to the MRI image-space and resample to 1mm isotropic resolution using the ANTS toolbox interface provided via Nipype \cite{gorgolewski2011nipype}. To increase the regularity and identifiability of MR structures, we apply contrast-limited adaptive histogram normalization to the resampled MRI volumes, using kernel-size of 100~mm and clipping limit of 0.05~\cite{zuiderveld1994contrast}. For consistency, we convert the raw PET intensity values (counts) to SUV using known radiotracer dose, half-life, positron-fraction, elapsed time, and patient weight~\cite{byrd2016metrics}. Finally, we split our dataset into 40 whole-body PET/MRI training exams, 16 whole-body PET/MRI testing exams, and 20 independent pelvic PET/MRI testing exams where corresponding CT was available (discussed in Section~\ref{sec:quantification}). We make no explicit assumption of age, race, gender, or ailment, other than through the image characteristics of the acquired dataset. 

\subsection{Model Architecture}  \label{subsec:architecture}
By fiat, we choose a 3D residual UNet architecture that combines the well-studied
{
2D/3D UNet~\cite{cciccek20163d} with residual (skip) connections~\cite{alom2018recurrent}
}
and convolutional upsampling~(Fig.~\ref{fig:architecture}). In our implementation, we take a one-channel 3D MRI volume as input, and employ 3x3x3 convolutional kernels followed by 2x2x2 maxpooling in each layer of the encoder (channel dimensions:~[32,64,128,256,512]), and 3x3x3 convolutional upsampling kernels in each layer of the decoder (channel dimensions:~[256, 128, 64, 32]), ultimately resulting in a one-channel 3D output. This architecture can be adapted to multi-channel inputs (multi-contrast MRI) and outputs (multiple PET radiotracers and/or dose levels) by modifying the first and last layers of the network respectively.

Inference is performed by breaking large whole-body MRI volumes into smaller overlapping volumetric patches with dimensions divisible by 32 (e.g.~[128 x 128 x 128]mm, with 50\% overlap) prior to applying the 3D UNet, and taking the sample mean of the resulting outputs at each 3D grid position to assemble the full whole-body volume. While the aforementioned resampling ensures MRI is processed at nearly native resolution to allow recognition of fine structural details, the PET groundtruth is considerably upsampled, especially in the $z$ dimension. This can be remedied by resampling the predicted volumes to the native PET image-space and resolution, e.g.~prior to performing PET/MRI reconstruction~(Section~\ref{sec:quantification}).

\subsection{Learning}  \label{subsec:learning}
One of the primary challenges with domain translation of MRI to PET is maintaining high accuracy across the full dynamic range of PET. Although SUV-scaling does provide a more consistent and intuitive numerical range, we find that explicit control in the objective function is required to prevent smoothing over suitable minima. For example, the histogram distribution of a whole-body $^{18}$F-FDG-PET exam (Fig.~\ref{fig:PEThisto}) reveals that different tissues differ in amount of physiologic uptake.
For example, in the lungs, heart, and liver there is often increased activity between $[1,4]$ SUV, and in regions such as the bladder and brain the recorded SUV can be greater than 20. In particular, since we are interested in using the predictions of our model for PET quantification studies, we require high accuracy across \textit{all} relevant scales. This precludes the use of simple $p$-norm objective functions, such as the mean absolute error (MAE), that may be dominated by the high absolute or relative error in one or more histogram bins.

To address this, we minimize the balanced objective:
\begin{align}
    J_{\text{total}} = J + \lambda J_{\text{LOR}} \label{eq:Jtotal}
\end{align}
{
where $J_\text{LOR}$ represents a regularization function with parameter $\lambda$, and $J$ is a linear combination of absolute and relative errors across $B$ different histogram bins, expressed as:
}
\begin{align}
    J = \sum_{j=1}^{B} \frac{\alpha_j}{|h_j|} \sum \mathcal{E}[h_j] + \sum_{k=1}^B \frac{\beta_k}{|h_k|} \sum \frac{\mathcal{E}[h_k]}{y[h_k] + \epsilon} \label{eq:J}
\end{align}
where $\mathcal{E} = | F(x) - y |$ is the conventional voxel-wise absolute error, $x$ is the MRI input volume, $y$ is the groundtruth PET volume, and $F(x)$ represents the predicted synthetic sPET. In Equation~\ref{eq:J}, $h_j$ represents an indicator variable selecting the voxels belonging to bin $j$ of the $B$-bin histogram of $y$, and $\epsilon$ is chosen as 1e-3 to prevent overflow. The histogram bins (Fig.~\ref{fig:PEThisto}) and corresponding weights ($\bar{\alpha}=[1,1,1,1,0]$, $\bar{\beta}=[0,0,1,1,1]$) were chosen based on empirical observation to prevent domination of $J$ by high absolute errors in high-SUV regions or by high relative errors in low-SUV regions.
{
The intention of this flexible formulation with $\bar{\alpha}$ and $\bar{\beta}$ is to define a family of functionals that can be tailored to different patient datasets, PET tracers, and anatomic regions.
}

To further improve both the perceptual image quality and convergence, during training we integrate and compare the groundtruth PET $y$ and the predicted sPET $\hat{y} = F(x)$ along random angles using a projection operator $\mathcal{R}_{\theta,\phi}$, mimicking tomographic data collection in a uniform, isotropic attenuating media along hypothetical PET lines of response (LOR), as:
\begin{align}
    J_{\text{LOR}} = \| \mathcal{R}_{\theta,\phi} \cdot (F(x) - y)/{(y+\epsilon)} \|_2 \label{eq:JLOR}
\end{align}

In addition to tying together the performance of different tomographically-related voxels, $J_{\text{LOR}}$ measures the error in the coarse \textit{scale} of predictions~on a line-by-line basis. For example if a 3D image patch shows little to no activity, $\|\mathcal{R}_{\theta,\phi}y\|_2$ will be nearly zero, whereas a patch from a region with high uptake may yield either high- or low-valued $\|\mathcal{R}_{\theta,\phi}y\|_2$. This improves convergence
{
and combats overfitting
by supervising the spatial distribution of sPET
}
without explicit assumptions of patient anatomy.

For all results shown in this paper, we used the Adam optimizer with initial learning rate of 1e-4, weight decay of 1e-3, and effective batchsize of 16 [128x128x128]mm volumetric patches generated systematically (in a random order) from the aforementioned whole-body $^{18}$F-FDG-PET/MRI dataset.

To improve convergence during training, we defined a custom 3D image patch sampler that performs round-robin sampling of different PET/MRI phenotypes present in the training dataset. 
{
These phenotypes were determined by first cataloging all the volumetric patches in the training dataset and computing their intensity histograms. Using k-means clustering ($K=10$), we computed a semantic grouping of these histograms that defined the different PET/MRI phenotypes that were sampled cyclically during model training. 
}

\begin{figure*}[t]
    \centering
    \includegraphics[width=0.93\linewidth]{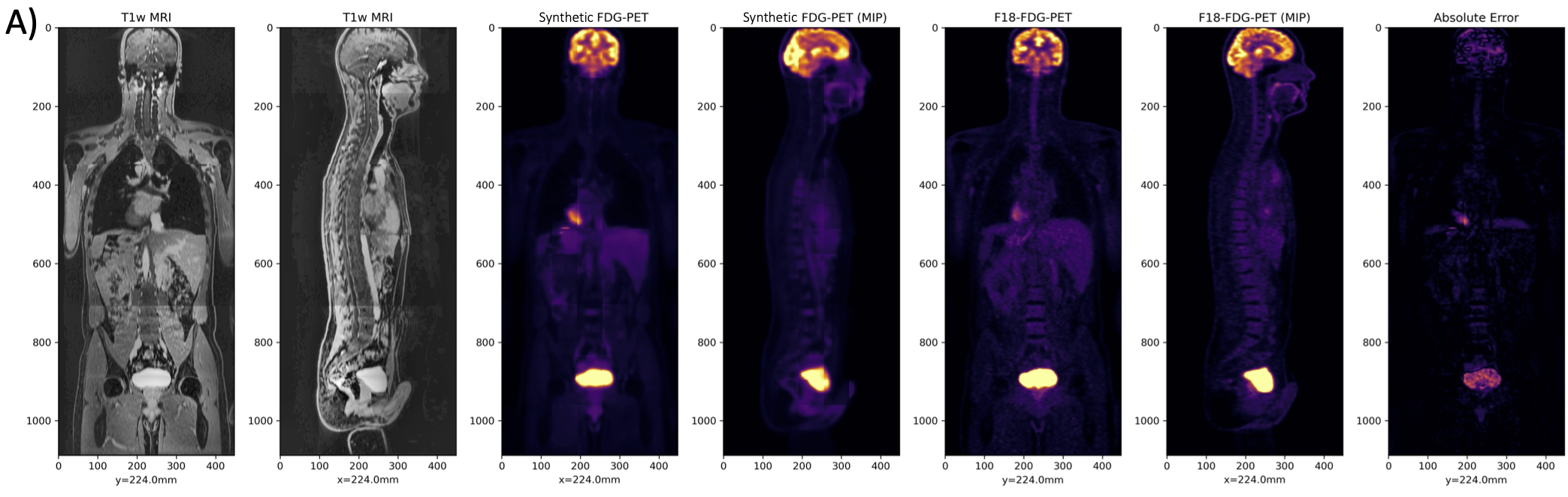} \\
    \includegraphics[width=0.93\linewidth]{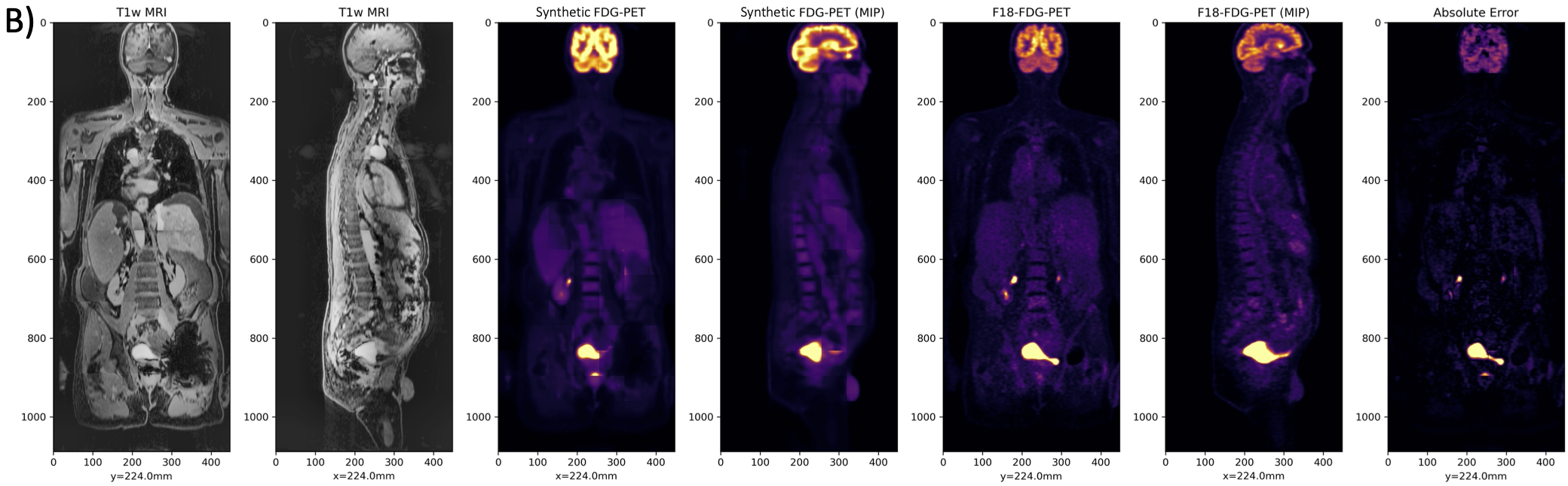} \\
    \includegraphics[width=0.93\linewidth]{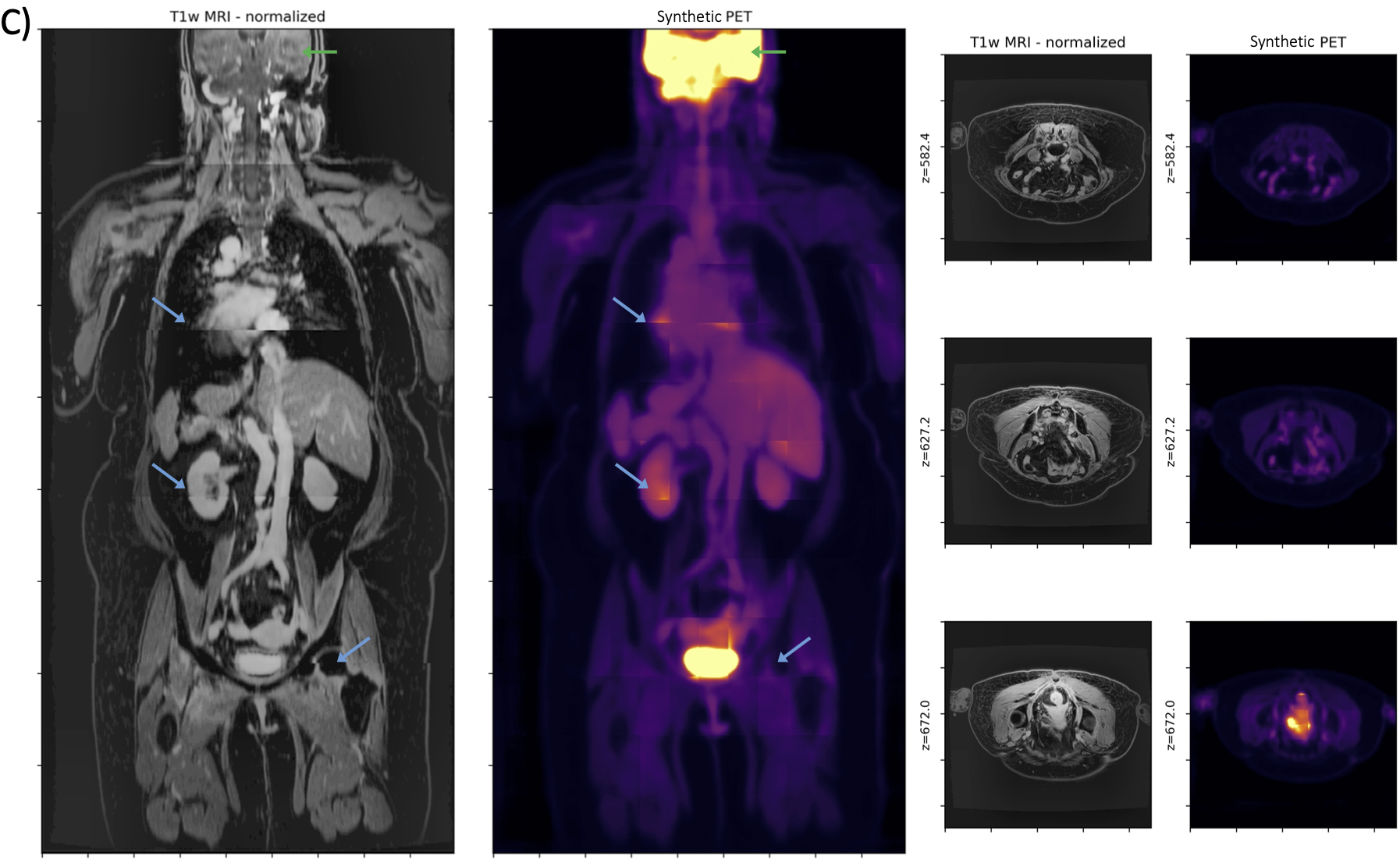}
    \caption{(A,B) A test-set evaluation of whole-body MR-based synthetic FDG-PET (sPET) in comparison to real $^{18}$F-FDG-PET/MRI. sPET mimics the typical physiologic uptake of FDG, showing high uptake in the brain and bladder as well as moderate uptake in liver, kidneys, heart and muscle.
    High relative error with the real PET data is expected in many regions where there is typically high physiologic variability between subjects (e.g. tumors, heart, bladder). While (A,B) represent patient exams from the intentionally withheld test set, (C) represents an exam from the additional validation set (with corresponding Pelvic PET/CT) exhibiting significant stitching artifacts (blue arrows) in the T1w-MRI between bed positions as well as loss in resolution in the head (green arrows). Various transverse slices in the abdomen are shown for comparison on the right of (C). Evaluation and inclusion of this exam in the validation cohort demonstrates that the proposed 3D UNet is able to recover reasonable FDG-uptake even in the presence of significant domain shift, a common issue when applying deep learning algorithms to clinical data acquired on a different scanner, or with different imaging protocols and image quality checks.}
    \label{fig:zdPET_results}
\end{figure*}

\subsection{Image Quality Metrics}  \label{subsec:errors}
We measure the quality of predicted sPET using quantitative error metrics, including the mean absolute error (MAE), mean relative absolute error (MRAE), and the 3D structural similarity index measure (SSIM). For each exam we compute MAE over all voxels $N$, as:
\begin{align}
    \text{MAE} = \frac{1}{N} \sum_{n} \|y_n - \hat{y}_n\|_1 , \label{eq:MAE}
\end{align}
while we compute MRAE only over voxels $K$ of at least 0.1 SUV, as:
\begin{align}
    \text{MRAE} = \frac{1}{K} \sum_{k} \frac{\|y_k - \hat{y}_k\|_1}{y_k} \label{eq:MRAE} .
\end{align}

The 3D-SSIM captures this information in a different way, accounting for differing scales and magnitudes through a measure of correlation within a 3D window, as:
\begin{align}
    {\displaystyle {\hbox{SSIM}}(x,y)={\frac {(2\mu _{x}\mu _{y}+c_{1})(2\sigma _{xy}+c_{2})}{(\mu _{x}^{2}+\mu _{y}^{2}+c_{1})(\sigma _{x}^{2}+\sigma _{y}^{2}+c_{2})}}} \label{eq:SSIM}
\end{align}
where $\mu_{x}$ and $\sigma^2_{x}$ represent the mean and variance of volume $x$, $\mu_{y}$ and $\sigma^2_{y}$ represent the mean and variance of volume $y$, $\sigma_{xy}$ represents the covariance of $x$ and $y$, and $c_*$ is chosen proportional to the dynamic range of pixel values~\cite{wang2003multiscale}.

\subsection{Results on Whole-Body $^{18}$F-FDG PET-MR Datasets}  \label{subsec:syntheticresults}
We find that prediction of synthetic FDG-PET, domain-translated from T1-weighted post-contrast MRI, works well despite the lack of salient tracer-specific or functional information in MRI~(Fig.~\ref{fig:zdPET_results}). Numerical results comparing the effect of different training objectives on test-set performance is shown in Table~\ref{tab:zdPET_results}. Qualitative analysis reveals that physiologic uptake is predicted accurately and reconstructed seamlessly throughout the body without obvious spatial artifacts, except in regions where we expect variable uptake (e.g.~heart, bladder). In the myocardium, for example, FDG-PET uptake depends on patient metabolism, which can vary across exams for even a single patient. Similarly, in the bladder PET uptake is often dependent on a patient's water consumption and timing of voiding~\cite{bach2012variation}.

\begin{table}[hbt!]
    \centering
    \resizebox{\columnwidth}{!}{
    \begin{tabular}{c|c|c|c}
        \textbf{Objective} &
        \textit{MAE}
        (Eq.~\ref{eq:MAE})
        & \textit{MRAE} 
        (Eq.~\ref{eq:MRAE})
        & \textit{3D-SSIM}
        (Eq.~\ref{eq:SSIM})
        \\
        \hline
        MAE (Eq.~\ref{eq:MAE}) & $0.090 \pm 0.021$ & $0.654 \pm .074$ & $0.473 \pm 0.073$ \\ 
        $J$ (Eq.~\ref{eq:J}) & $0.083 \pm 0.031$ & $0.487 \pm .145$ & $0.863 \pm 0.089$ \\ 
        $J_\text{total}$  (Eq.~\ref{eq:Jtotal}) & $\textbf{0.066} \pm 0.026$ & $\textbf{0.369} \pm .092$ & $\textbf{0.938} \pm 0.060$  
    \end{tabular}
    }
    \caption{Test-set performance $\pm$ 1 stdev using different supervised MR-to-PET domain translation training objectives}
    \label{tab:zdPET_results}
    \vspace{-2mm}
\end{table}

The MAE and MRAE results show that incorporation of both balanced histogram losses and and tomographic projection-based losses can significantly reduce the quantitative error in prediction of synthetic PET from MRI. The SSIM results show that this reduction in error boosts the image quality of the synthetic PET image relative to the real PET image. The inclusion of SSIM is important to assess the realness of synthetic PET, in lieu of reporting MAE and MRAE within different organs and anatomical structures.

\section{PET Quantification using Synthetic PET} \label{sec:quantification}
PET/MRI quantification is important for establishing the accuracy and reproducibility of PET reconstructions when the photon attenuation maps are inferred entirely from MRI.
As the error in PET/MRI reconstruction is composed of errors involving prediction of the attenuation map and errors involving the reconstruction (e.g.~choice of objective function), a standard approach is to measure the compound effect caused by the AC map by directly comparing PET volumes reconstructed with MRAC and CTAC voxel-wise and regionally~\cite{hamdi2021evaluation,catana2021path}.

Specifically, we evaluate the applicability of our MR-derived synthetic PET imagery for algorithm development by replicating a MRAC vs CTAC PET SUV quantification task using synthetic PET data in place of real list-mode PET data. To achieve this we forward project synthetic PET data into sinogram space using vendor-provided software that incorporates scanner geometry, detector response, and normalization.

\subsection{Reconstruction Model and Parameters} \label{subsec:reconparameters}

For time-of-flight PET (ToF-PET), the measured sinogram data is modelled within the forward model as~\cite{wadhwa2021pet}:
\begin{align}
    y_{pt} = A_{pt} x + b_{pt} \label{eq:PET_forward_model}
\end{align}
where $y_{pt}$ represents the ToF projection data measured by the scanner, $x$ is the PET image to be found, and the system matrix $A$ models the probability of an event emitted in voxel $m$ to be detected by detector pair $p$ within the signed timing bin number $t$, summarizing the attenuation of the media along PET LoR, patient-scanner geometry, and detector efficiencies. $b_{pt}$ corresponds to the background counts of the timing bin $t$ and detector pair $p$.
For this model, a basic reconstruction approach is to solve the optimization problem:
\begin{align}
    \hat{x} = \min_x || A x - b ||_2 + \mathcal{R}(x)
\end{align}
where $\mathcal{R}$ is a regularization function (e.g. total-variation). In practice, vendor-provided ordered-subset expectation-maximization (OSEM) or ToF-OSEM with point-spread-function (PSF) modeling are used for clinical imaging~\cite{wadhwa2021pet,rogasch2020reconstructed}. In our experiments, we utilize the following clinical image reconstruction parameters for the GE Signa PET/MRI:
\begin{table}[hbt!]
    \vspace{-2mm}
    \centering
    \begin{tabular}{|c|c|}
        \hline
        \textbf{Parameter} & \textbf{Value} \\
        \hline
        Objective Function & ToF-OSEM-PSF \\
        \hline
        Subsets &  28 \\
        \hline
        Iterations & 2 \\
        \hline
        Transverse Filter (FWHM) & 2.0~mm \\
        \hline
        Axial Filter (FWHM) & 4.0~mm \\
        \hline
        Transverse Field-of-View (FOV) & 600~mm\\
        \hline
        Transverse matrix size & [256,256] \\
        \hline
        \# Projection Angles / Views & 180 \\
        \hline
    \end{tabular}
    \caption{Summary of PET Reconstruction Hyperparameters.} \label{tab:reconparams}
    \vspace{-7mm}
\end{table}

\subsection{Synthetic Sinogram Generation and Lesion Insertion} \label{subsec:syntheticsinogram}
For a given system matrix $A$, a reconstructed PET image $x$ can be projected into the sinogram domain by applying the forward model (Eq.~\ref{eq:PET_forward_model}) to yield $y_\text{simulated}$. The forward projection tool provided with the Duetto toolbox (v02.03, GE Healthcare) performs this operation on a synthetic volume of dimension equal to the reconstructed volume, to generate a synthetic lesion sinogram that is added to the sinogram corresponding to $x$. Image reconstruction can then be performed on this ``lesion-inserted'' sinogram, as if it were the real sinogram, using a variety of methods (e.g.~ToF, PSF, regularization).

\begin{figure}[hbt!]
    \vspace{-2mm}
    \centering
    \includegraphics[width=\linewidth]{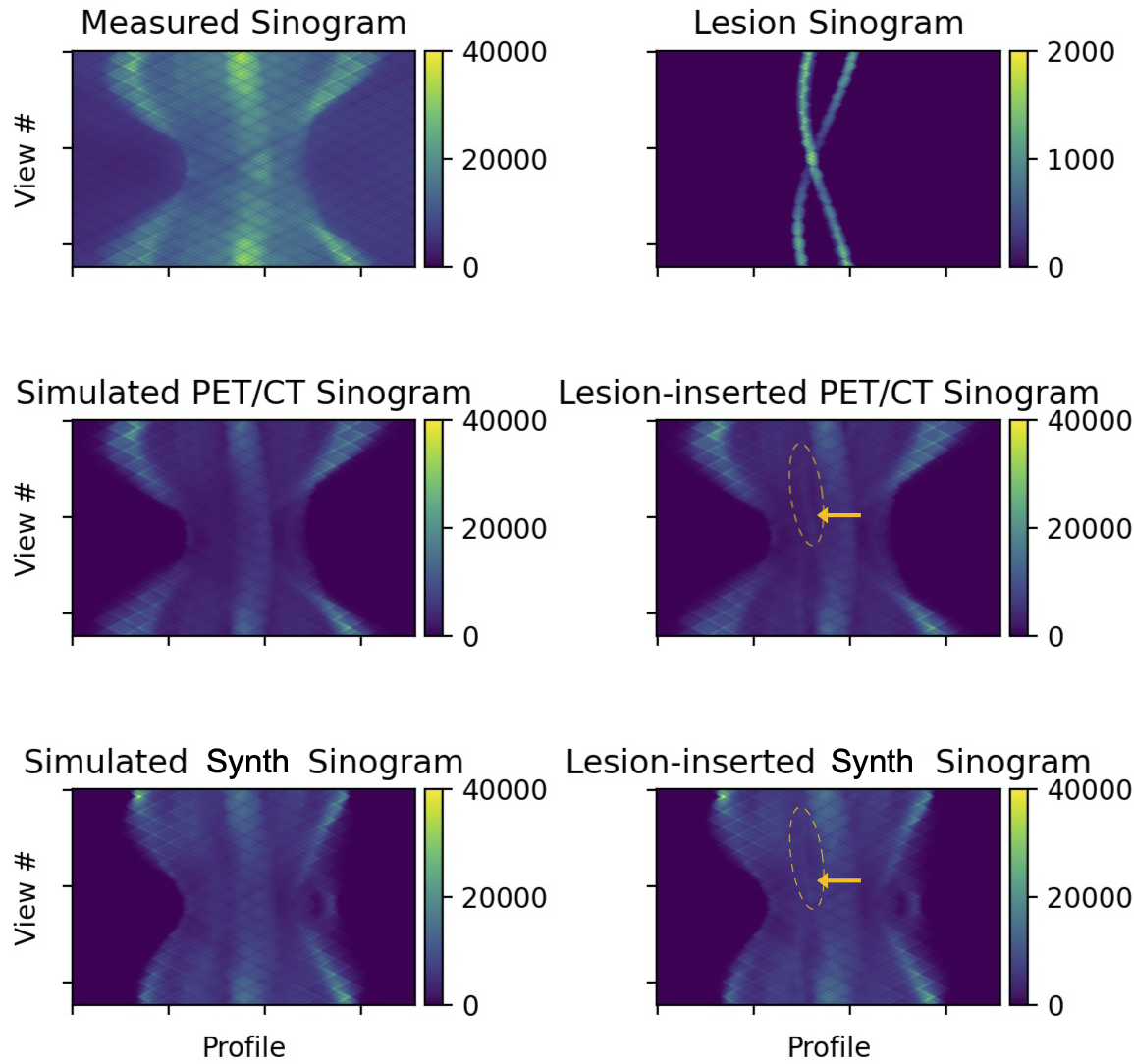}
    \vspace{-6mm}
    \caption{Measured and simulated sinograms representing different PET sources with corresponding synthetically-inserted lesion sinograms. The annotation (yellow arrow) highlights a region affected by lesion insertion.
    }
    \vspace{-4mm}
    \label{fig:sinogram}
\end{figure}

We exploit this mechanism to generate synthetic PET sinogram data from domain-translated synthetic PET imagery. However, as Duetto currently does not incorporate scatter simulation, we perform reconstructions with scatter estimation and correction turned off. As this introduces an additional discrepancy between real PET and synthetic PET reconstructions, in both cases we start by forward projecting a 3D ``source'' volume $x_\text{source}$ to yield a simulated ToF-sinogram that is subsequently inserted with synthetic lesions (Fig.~\ref{fig:sinogram}).

\begin{figure*}[b]
    \vspace{-4mm}
    \centering
    \includegraphics[width=0.9\linewidth]{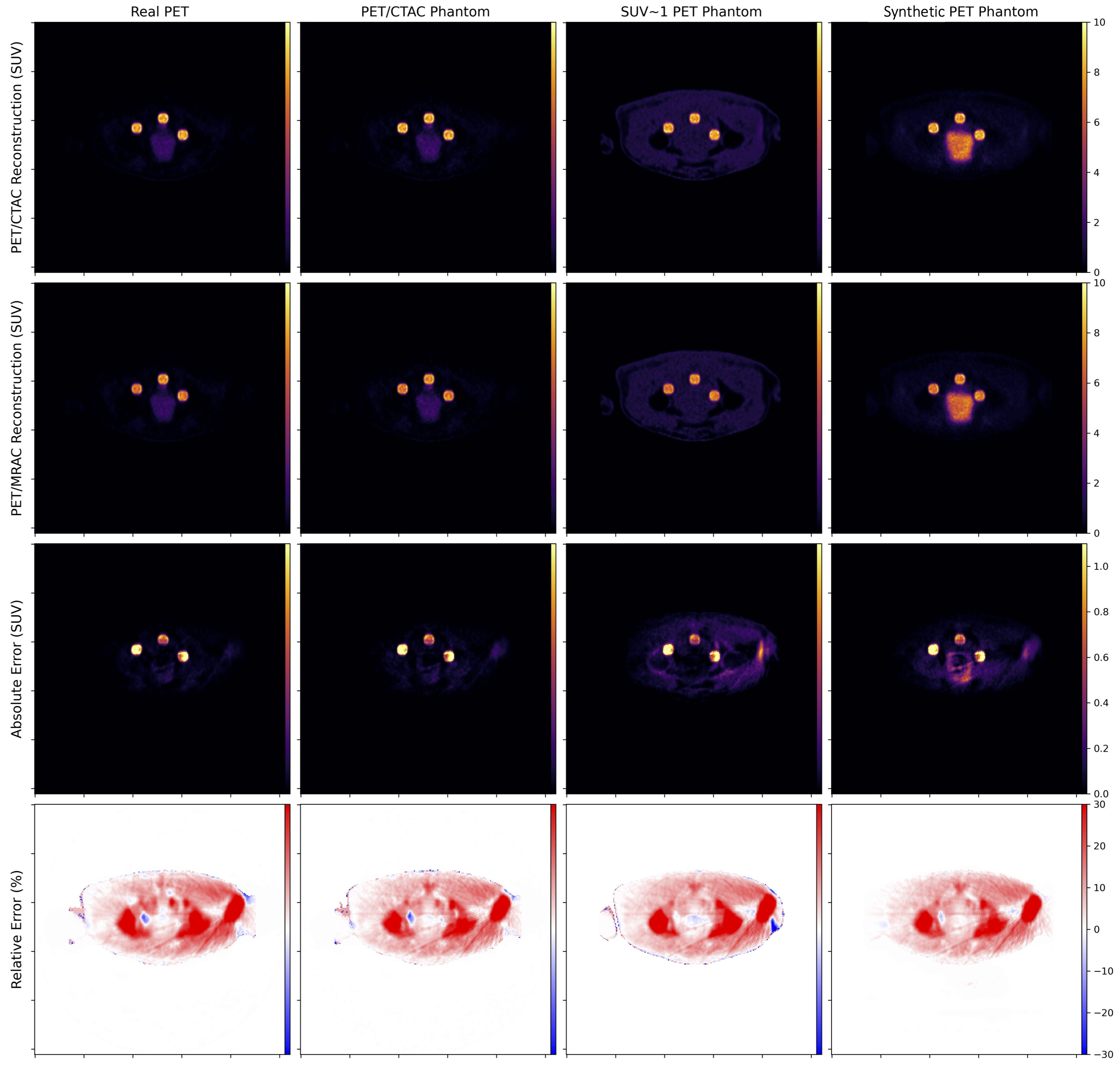}
        \vspace{-4mm}
    \caption{Example evaluation of synthetically inserted lesions into 3D reconstructions using various PET data sources (anterior is superior in our presentations). For PET data source (columns), we compute a reconstruction using CTAC and MRAC, and compute the absolute and relative errors for each slice. Shown here is a single slice from a single patient with contributions from 3 synthetically inserted lesions. The error in the sPET prediction is considerably lower than using the phantom with SUV\tildee1, and has a similar distribution to using real PET data.}
    \label{fig:quantification}
\end{figure*}

\subsection{Quantification Experiment Summary} \label{subsec:quantificationexp}
The pelvic CTAC vs MRAC FDG-PET reconstruction and SUV quantification experiment can be summarized as follows:
\begin{enumerate}
    \item Forward project $x_\text{source}$ using a registered CT-based attenuation map to yield sinogram $y_\text{simulated}$. To evaluate the applicability of different synthetic PET sources for this pipeline, we choose $x_\text{source}$ as:
    \begin{itemize}
        \item \textbf{Real PET} $x_\text{real}$: the true patient activity distribution, corresponding to measured patient sinogram $y_\text{real}$.
        \item \textbf{Reconstructed patient phantom} $x_\text{live}$: a PET/CT image volume, reconstructed from measured PET sinogram data with a CT-based attenuation map.
        \item \textbf{Uniform SUV\tildee1} $x_\text{uniform}$: we threshold a T1-weighted postcontrast MRI volume to define a body-mask filled with activity corresponding to SUV~1.
        \item \textbf{Synthetic sPET} $x_\text{synthetic}$: a synthetic PET volume generated from a T1-weighted postcontrast MRI using the aforementioned 3D UNet.
    \end{itemize}

    \item Forward project synthetic lesions specified by a 3D volume $x_\text{lesion}$ to yield $y_\text{lesion-simulated}$. In our experiments, a board-certified radiologist annotated 4 sites for lesion insertion in each pelvic MR exam: in the acetabulum, sacrum, rectum, and lymph nodes. These locations were specifically identified to challenge the ability of MR-based reconstruction to reproduce activity surrounded by soft-tissue and bone. For each location, a spherical lesion with diameter 12mm and activity corresponding to SUV~8 was added to a zero-filled $x_\text{lesion}$ volume.
    \item Reconstruct lesion-inserted sinograms using vendor-provided CT-based attenuation correction (CTAC) and MR-based attenuation correction (MRAC) methods (with parameters specified in Section~\ref{subsec:reconparameters}), resulting in PET images $\hat{x}_\text{CT}$ and $\hat{x}_\text{MR}$ respectively for each $x_\text{source}$.
    
    \item Evaluate voxel-wise and regional absolute and relative error between $\hat{x}_\text{CT}$ and $\hat{x}_\text{MR}$ in each lesion volume of interest (VOI) for each $x_\text{source}$ for each exam. Evaluation within each VOI can also provide a quantiative measure of accuracy, since the activity was synthetically inserted.
\end{enumerate}

\begin{table*}[b]
    \centering
    \begin{tabular}{c|c|c|c|c|c|c|}
        & \multicolumn{6}{|c|}{\textit{Average Difference in Mean-SUV between PET/CTAC and PET/MRAC Reconstructions in}} \\
        \textbf{PET Data Source} & \textit{All Lesions} & \textit{Lesion (Acetabulum)} & \textit{Lesion (Sacrum)} & \textit{Lesion (Rectum)} & \textit{Lesion (Lymph)} & \textit{Background}\\
        \hline
        $x_\text{real}$ &  0.474 &       0.630 &   0.627 &   0.271 &  0.410 &  0.011 \\
        \hdashline
        $x_\text{live}$ &  0.485 &       0.644 &   0.647 &   0.276 &  0.422 &  0.012 \\
        $x_\text{uniform}$        &  0.505 &       0.694 &   0.688 &   0.262 &  0.429 &  0.033 \\
        $x_\text{synthetic}$ &  0.498 &       0.658 &   0.666 &   0.277 &  0.439 &  0.024 \\
    \end{tabular}
    \begin{tabular}{c|c|c|c|c|c|c|}
        & \multicolumn{6}{|c|}{\textit{Average Difference in Max-SUV between PET/CTAC and PET/MRAC Reconstructions in}} \\
        \textbf{PET Data Source} & \textit{All Lesions} & \textit{Lesion (Acetabulum)} & \textit{Lesion (Sacrum)} & \textit{Lesion (Rectum)} & \textit{Lesion (Lymph)} & \textit{Background}\\
        \hline
        $x_\text{real}$ &  0.524 &       0.985 &   1.015 &   0.353 &  0.606 &  0.539 \\
        \hdashline
        $x_\text{live}$ &  0.484 &       0.995 &   1.043 &   0.348 &  0.608 &  0.545 \\
        $x_\text{uniform}$  &  0.586 &       1.017 &   1.053 &   0.305 &  0.602 &  0.586 \\
        $x_\text{synthetic}$ &  0.512 &       0.991 &   1.022 &   0.328 &  0.588 &  0.476 \\
    \end{tabular}
    \begin{tabular}{c|c|c|c|c|c|c|}
        & \multicolumn{6}{|c|}{\textit{Average Difference in Peak-SUV between PET/CTAC and PET/MRAC Reconstructions in}} \\
        \textbf{PET Data Source} & \textit{All Lesions} & \textit{Lesion (Acetabulum)} & \textit{Lesion (Sacrum)} & \textit{Lesion (Rectum)} & \textit{Lesion (Lymph)} & \textit{Background}\\
        \hline
        $x_\text{real}$ &  0.461 &       0.912 &   0.945 &   0.310 &  0.600 &  0.488 \\
        \hdashline
        $x_\text{live}$ &  0.371 &       0.929 &   0.970 &   0.302 &  0.592 &  0.457 \\
        $x_\text{uniform}$  &  0.540 &       0.979 &   1.006 &   0.302 &  0.607 &  0.540 \\
        $x_\text{synthetic}$ &  0.484 &       0.943 &   0.976 &   0.335 &  0.555 &  0.406 \\
    \end{tabular}
    \caption{Absolute errors between CTAC- and MRAC-based reconstructions for various real and synthetic PET data sources.}
    \label{tab:quant_results_delta}
    \vspace{-2mm}
\end{table*}

\begin{table*}[b] 
    \centering
    \begin{tabular}{c|c|c|c|c|c|c|}
        & \multicolumn{4}{|c|}{\textit{Percent Error in Mean-SUV compared to Baseline ($x_\text{live}$) in}} \\
        \textbf{PET Data Source} & \textit{Lesion (Acetabulum)} & \textit{Lesion (Sacrum)} & \textit{Lesion (Rectum)} & \textit{Lesion (Lymph)} \\
        \hline
        $x_\text{real}$      &  2.48 $\pm$ 2.10  &  7.56 $\pm$ 8.50  &    3.85 $\pm$ 4.22  &  4.06 $\pm$ 3.16  \\
        \hdashline
        $x_\text{uniform} $  &  7.75 $\pm$ 2.20  &  7.34 $\pm$ 4.57  &  22.09 $\pm$ 2.86  &  \textbf{2.47 }$\pm$ 1.46  \\
        $x_\text{synthetic}$ &  \textbf{2.94} $\pm$ 1.50  &  \textbf{5.54} $\pm$ 3.80  &  \textbf{15.88} $\pm$ 2.07  &  6.17 $\pm$ 1.59   \\
    \end{tabular}
    \begin{tabular}{c|c|c|c|c|c|c|}
        & \multicolumn{4}{|c|}{\textit{Percent Error in Max-SUV compared to Baseline ($x_\text{live}$) in}} \\
        \textbf{PET Data Source} & \textit{Lesion (Acetabulum)} & \textit{Lesion (Sacrum)} & \textit{Lesion (Rectum)} & \textit{Lesion (Lymph)} \\
        \hline
        $x_\text{real}$      &  7.57 $\pm$ 6.48  &    8.23 $\pm$ 4.94  &   80.61 $\pm$ 14.41  &  16.16 $\pm$ 1.58  \\
        \hdashline
        $x_\text{uniform} $  &  8.56 $\pm$ 6.46  &    \textbf{9.84} $\pm$ 7.05  &   \textbf{66.07} $\pm$ 14.54  &  17.74 $\pm$ 1.31 \\
        $x_\text{synthetic}$ &  \textbf{3.42} $\pm$ 3.49  &   13.53 $\pm$ 2.14  &  121.53 $\pm$ 15.78  &  \textbf{12.88} $\pm$ 1.78  \\
    \end{tabular}
    \begin{tabular}{c|c|c|c|c|c|c|}
        & \multicolumn{4}{|c|}{\textit{Percent Error in Peak-SUV compared to Baseline ($x_\text{live}$) in}} \\
        \textbf{PET Data Source} & \textit{Lesion (Acetabulum)} & \textit{Lesion (Sacrum)} & \textit{Lesion (Rectum)} & \textit{Lesion (Lymph)} \\
        \hline
        $x_\text{real}$      &  2.56 $\pm$ 2.07  &  15.96 $\pm$ 18.71  &   29.01 $\pm$ 58.60  &  5.28 $\pm$ 3.82  \\
        \hdashline
        $x_\text{uniform} $  &  2.73 $\pm$ 0.81  &  19.56 $\pm$ 29.70  &   \textbf{34.53} $\pm$ 40.99  &  \textbf{2.17 }$\pm$ 2.42  \\
        $x_\text{synthetic}$ &  \textbf{1.98} $\pm$ 1.48  &   \textbf{9.60} $\pm$ 11.61  &  92.30 $\pm$ 19.34  &  3.20 $\pm$ 1.93  \\
    \end{tabular}
    
    \caption{Deviation in predicted quantification error (lower is better) for various real and synthetic PET data sources.\\ $\pm$~indicates one standard deviation.}
    \label{tab:quant_results_deviation}
    \vspace{-2mm}
\end{table*}
In particular, we evaluate the ability of different synthetic sinograms (corresponding to a choice of $x_\text{source}$) to reproduce the CTAC vs MRAC ``quantification error'' $\Delta_\text{quant}$, normally estimated using real measured PET sinogram data. We quantify this by computing and comparing \textit{deviation of error} in mean-, max-, and peak-SUV between $\hat{x}_\text{CT}$ and $\hat{x}_\text{MR}$ for each $x_\text{source}$ in each VOI compared to using real PET sinogram data. That is, for each $x_\text{source}$ we compute:
\vspace{-2mm}
\begin{align}
    \Delta_\text{quant,source} &= 
    \|
    \texttt{quant}\big( \hat{x}_\text{CT}[\text{V}] \big)
    -
    \texttt{quant}\big( \hat{x}_\text{MR}[\text{V}] \big)
    \Big)
    \|_1
\end{align}
\vspace{-4mm}
\begin{align}
    \delta_\text{quant, source} &= | \Delta_\text{quant, true} - \Delta_\text{quant, source}|
\end{align}
\begin{align}
    \gamma_\text{quant, source} &= \frac{|\Delta_\text{quant, true} - \Delta_\text{quant, source}|}{ \Delta_\text{quant, true}} \times 100\%
\end{align}
where $\text{V}$ represents an indicator function for voxels in a VOI, $\texttt{quant}$ represents the mean-, max-, or peak-SUV computation in a VOI, we take $\Delta_\text{true}$ as the corresponding mean-, max-, or peak-SUV quantification error computed using the reconstruction patient phantom $x_\text{live}$ as the source, and $\delta$ and $\gamma$ represent absolute and relative quantification error respectively. To benchmark systematic error arising from the reconstruction and re-projection in the experimental procedure, we also compare to the quantification error arising from using measured sinogram data $y_\text{real}$ corresponding to the true patient distribution $x_\text{real}$ (i.e.~following the standard approach in~\cite{catana2021path}).

For each patient exam we select 5 different VOIs: lesion voxels corresponding to the four annotated regions (acetabulum, sacrum, rectum, and lymph), and ``background'', representing all non-zero voxels outside of the synthetic lesions. Quantification error is computed for each VOI by comparing mean-, max-, and peak-SUV between CTAC-based and MRAC-based reconstructions. Subsequently we compare the quantification error predicted by each PET data source to that predicted by the aforementioned reconstructed PET/CTAC live phantom. The absolute error is quantified for the background pixels, but the relative error is not since many voxels are devoid of any activity, positively skewing (overestimating) the mean relative error computation. In lieu of individual regions within the pelvis, the relative error in background voxels is better evaluated qualitatively by comparing slices in the transverse plane (Fig.~\ref{fig:quantification}).

\subsection{Results on Pelvic $^{18}$F-FDG PET/MR/CT Datasets}
\begin{figure}[hbt!]
    \centering
    \includegraphics[width=\linewidth]{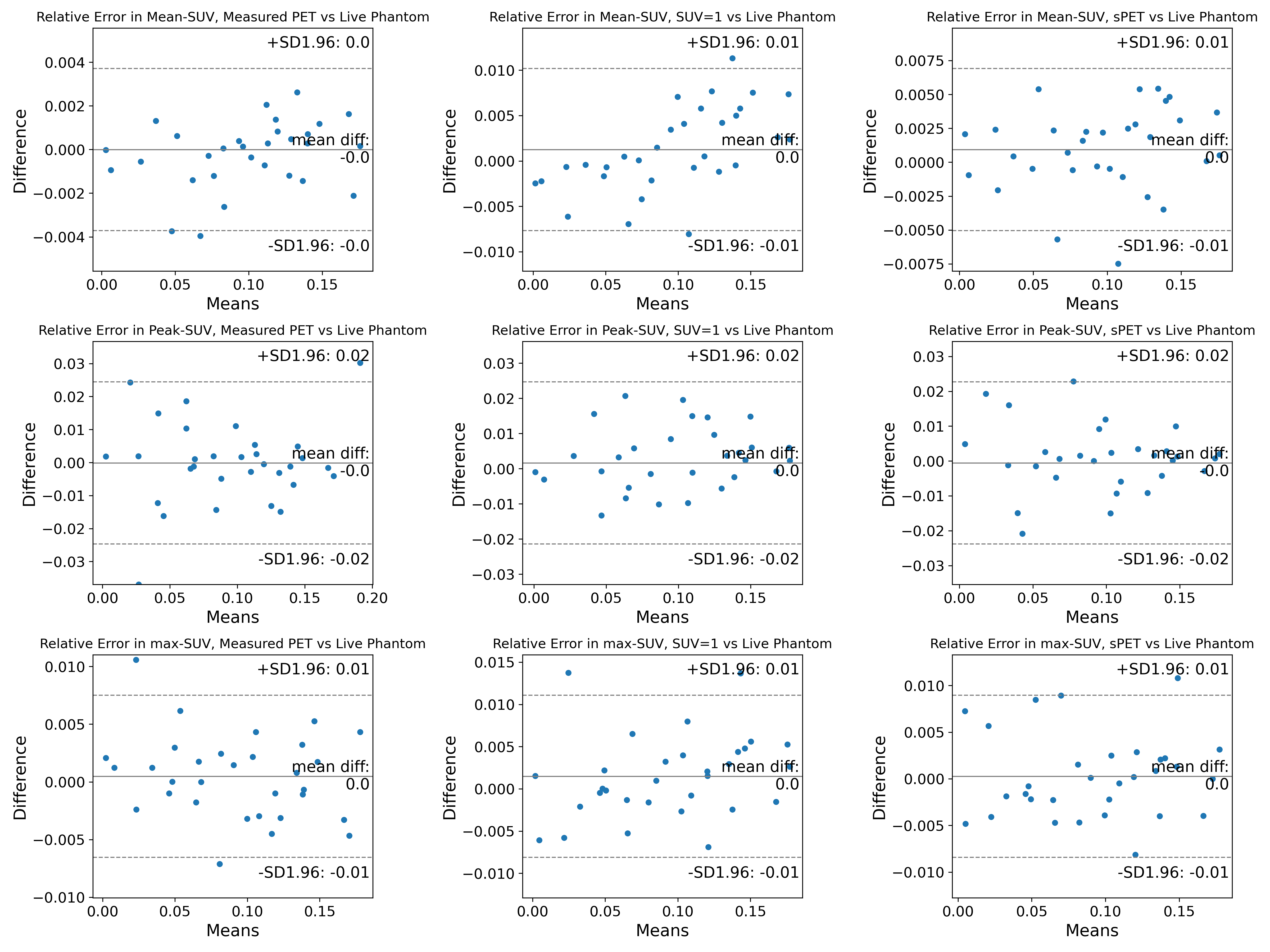}
    \caption{Bland-Altman plots that compare the CTAC-vs-MRAC \textit{Absolute Error} computed by the various types of phantoms and the Live Phantom. Each column represents a different synthetic PET phantom. Each row represents the error with respect to each metric of interest (mean-SUV, peak-SUV, or max-SUV). }
    \label{fig:BlandAltman_B}
\end{figure}

Numerical results presented in Tables~\ref{tab:quant_results_delta}-\ref{tab:quant_results_deviation} indicate that domain-translated MR-based synthetic PET (sPET) can achieve low absolute and relative deviation in quantification error compared to the quantification error predicted by the live PET/CTAC phantom source for synthetically-inserted pelvic lesions. 
Table~\ref{tab:quant_results_delta} shows that sPET-based evaluation to compare CTAC and MRAC-based reconstruction achieves SUV errors that were very similar to the measured PET-based evaluation across inserted lesions and in the background.
The percent quantification errors in Table~\ref{tab:quant_results_deviation} shows that sPET-based evaluation was more similar to measured PET-based evaluation than uniform SUV\tildee1-based evaluation, outperforming for mean-SUV evaluations across lesion types.
This suggests the applicability of synthetic sPET as a suitable replacement for real measured PET in PET-SUV quantification tasks. 
{
In the supplementary material (Figure~\ref{fig:BlandAltman_A} and~\ref{fig:BlandAltman_B}), we provide Bland-Altman plots that compare the CTAC-vs-MRAC error computed by the various types of phantoms and the Live Phantom. Each column represents a different synthetic PET phantom. Each row represents a different error metric (absolute error or relative error in mean-SUV, peak-SUV, or max-SUV). This analysis shows no significant differences between sPET and measured PET using the aforementioned figures of merit. 
}

\section{Discussion}
Overall, we have shown that MR-derived synthetic FDG-PET accurately captures the background physiologic distribution of PET imagery, creating 
{
images with realistic spatial distributions,
}
and that it can be combined with synthetic lesion insertion to provide data for evaluation of PET quantification methods.
However, the main limitation we observed is that it is smoother than corresponding full-dose imagery, perhaps due to the implicit regularization properties of convolutional networks (e.g.~exploited by DIP techniques~\cite{gong2018pet}). While this is desirable for enhancing low-dose or noisy PET/MRI, it is not entirely beneficial for our application due to mismatches in the intensity distribution used in the quantification experiment.

This is a good opportunity for future works that make use of generative adversarial networks (GANs), which may seek to better match the statistical distribution of sPET and PET
{
to increase its realism,
rather than simply regressing by value.
Note that a pure GAN approach based on noise vectors is not valid here because it may not provide anatomic conformity between the MRI and synthetic PET image, which is important to maintain for PET/MRI reconstruction algorithm research. Instead, adversarial losses may be added to the existing approach to increase realism and to help
}
reduce artifacts in the regions where variable uptake is expected, or where patch-based inference lacks sufficient context to prevent gridding or stitching artifacts
{
(although the effect of these artifacts is often reduced after forward projection to sinogram space).
In this respect, the physics-based tomographic LOR loss utilized in this paper not only works to increase the realism of synthetic PET but also improves its quantitative accuracy.
}

{
We believe that such physics-based approaches are crucial for the development of quantiative imaging and dataset generation techniques based on neural networks. While the tomographic LOR loss used in this work improves the quantitative error rates and qualitative realism (partially captured by SSIM) associated with sPET, advanced physics-based modeling could further improve both the realism as well as the applicability of the developed approach to more PET/MRI systems, e.g.~by utilizing their system matrix to optimize directly in the singoram domain, or measure congruence after image reconstruction.
}

Results from the 
{
downstream PET SUV
}
quantification experiment indicate that sPET can serve as an adequate surrogate for real data in a MRAC vs CTAC quantification experiments. This experiment also indicates that the PET background distribution does not significantly impact quantification performance when using synthetically-inserted lesions and without any scatter and randoms simulation.
Thus, further investigation of a more complete reconstruction is required to determine whether the PET background distribution affects quantification for real lesions. Based on the realistic appearance of sPET, we believe it will be an important tool in evaluations when accurate background distribution is required.

Although in some cases the estimate based on sPET underestimates the benchmark (reconstructed patient phantom) error, the strong agreement over a number of exams ($N=20$) and lesions indicates that sPET may be used as a qualification method when a large number of exams is required. This is precisely the domain for which sPET was designed, as large MRI/CT databases can be retrospectively utilized to establish a large synthetic PET/MRAC/CTAC dataset for scanner and algorithm qualification. Interestingly, in many cases the quantification error predicted by the synthetic sPET phantom more closely matches the quantification error predicted using real PET sinogram data, compared to even that of the live PET/CT phantom.
However, for most VOIs and phantom types, the deviation in quantification error is minimal. Here, synthetic sPET has an advantage over the uniform SUV\tildee1 phantom, because it a represents a realistic, anthropomorphic PET uptake pattern.

One limitation of our methodology is that it does not directly model noise or count statistics associated with PET data collection, which has been shown to impact the performance of PET reconstruction algorithms~\cite{tong2010noise,lodge2012noise, rahmim2013noise}. To address this, we note that the MR-based synthetic PET images proposed in this paper can be treated simply as an ideal source volume and, thus, readily combined with Monte Carlo PET simulators such as with GATE~\cite{jan2004gate}, SIMSPET~\cite{guerin2008realistic}, or SimPET~\cite{paredes2021simpet}. An alternative data-driven approach to address this issue may be to utilize a adversarial training, which can increase the realism of synthetic PET, thereby indirectly capturing statistical noise properties of PET acquisition in the image-domain.

\section{Future Work}
{
Detailing the generation of synthetic PET from 3D MRI and, importantly, demonstrating its utilize in downstream qualification research, opens the path to new research directions that can enable us to study new PET image reconstruction algorithms that can address important clinical questions. For example, virtual PET clinics have been previously proposed as a technique to evaluate PET detector systems and patient studies in a virtual simulation environment~\cite{abadi2020virtual}. This could also be extended to address 4D PET/CT and PET/MRI modalities~\cite{tsoumpas2013modeling}, enabling new approaches to diagnose cancers, such as the identification of recurrent gliomas using FET PET~\cite{debus2018impact,lohmeier2019quantitative}. In another vein, synthetic PET can also be used to directly improve image reconstruction algorithms themselves, e.g.~by generation of an deep learning prior image that can help regularize PET image reconstruction~\cite{rajagopal2021enhanced}. These applications provide a strong motivation for future work in curating large databases of PET/MRI with multiple MRI contrasts and PET radiotracer images, which could mirror and complement the impact of other synthetic MRI~\cite{hagiwara2016contrast,ji2020synthetic}. In this respect, the methods developed in the paper provide the framework and context necessary for such development.
}

\section{Conclusion}
In conclusion, we have demonstrated a method using deep learning to generate realistic, synthetic whole-body PET data from MRI, and that it is a suitable substitute for real PET data in a reconstruction evaluation task. The synthetic PET data, which mimics physiologic tracer distribution, can be combined with synthetic lesion insertion to mimic abnormal regions of high update. We demonstrated its equivalent performance to real PET data for comparing CTAC and MRAC for PET reconstruction, and believe this result combined with the apparent realism of the synthetic images will make this method broadly applicable for evaluating the robustness of PET/MRI reconstructions and component techniques, including attenuation correction, scatter correction, and MR-guided reconstruction algorithms, using large and diverse patient datasets.

All source code for this paper, including synthetic PET training code, PET reconstruction wrappers, and quantification experiments, is available freely at:
\begin{center}
{\color{blue}{\href{https://gitlab.com/abhe/SyntheticPET-TRPMS22}{https://gitlab.com/abhe/SyntheticPET-TRPMS22}}}
\end{center}

\section*{Acknowledgements}
This work was supported by NIH/NIBIB Grant \#F32EB030411 and NIH/NCI Grant \#R01CA212148. All authors declare that they have no known conflicts of interest in terms of competing financial interests or personal relationships that could have an influence or are relevant to the work reported in this paper.

\bibliographystyle{ieeetr}
\bibliography{refs}

\newpage
\onecolumn
\section*{Supplementary Material}
\begin{figure}[hbt!]
    \centering
    \includegraphics[width=\linewidth]{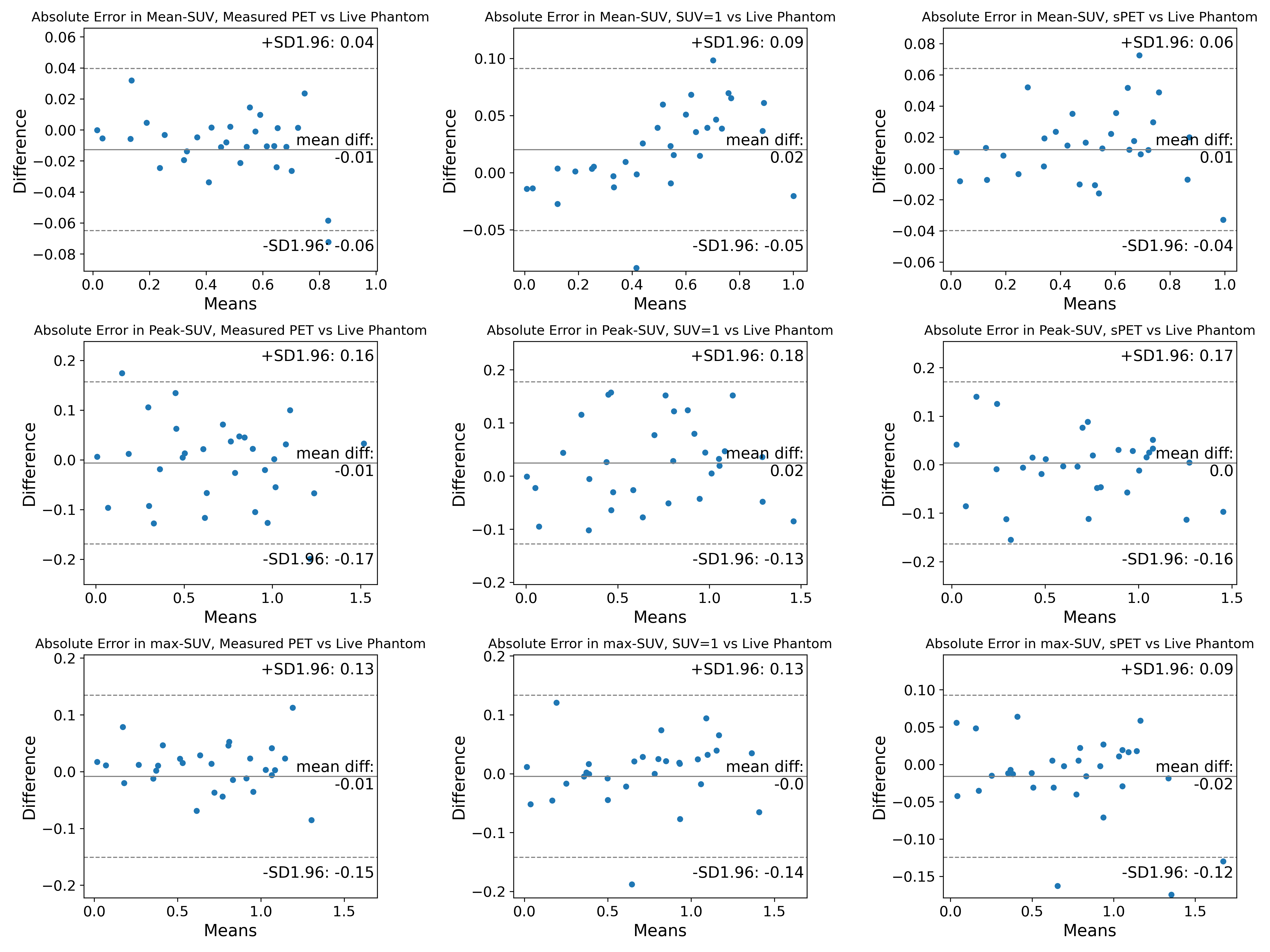}
    \caption{Bland-Altman plots that compare the CTAC-vs-MRAC \textit{Absolute Error} computed by the various types of phantoms and the Live Phantom. Each column represents a different synthetic PET phantom. Each row represents the error with respect to each metric of interest (mean-SUV, peak-SUV, or max-SUV). }
    \label{fig:BlandAltman_A}
\end{figure}

\end{document}